\documentclass[jcp,preprint,a4]{revtex4-1}
\usepackage[margin=3cm]{geometry}
\usepackage[utf8]{inputenc}
\usepackage{amsmath}
\usepackage{amsfonts}
\usepackage[usesnames,dvipsnames]{xcolor}
\usepackage{pifont}
\usepackage{indentfirst}
\usepackage{graphicx}
\usepackage{microtype}
\usepackage{subcaption}

\begin{document}

\title{Stochastic Electrochemical Kinetics}
\date{\today}
\author{Otávio Beruski}
\affiliation{Instituto de Química de São Carlos, Universidade de São Paulo,
400 Av. Trabalhador Sao-carlense, São Carlos, São Paulo, Brazil}
\email{oberuski@iqsc.usp.br}

\begin{abstract}
    A new tool for modeling electrochemical kinetics is presented. An
extension of the Stochastic Simulation Algorithm framework to electrochemical
systems is proposed. The physical justifications and constraints for the
derivation of a chemical master equation are provided and discussed. The
electrochemical driving forces are included in the mathematical framework, and
equations are provided for the associated electric responses. The
implementation for potentiostatic and galvanostatic systems is presented,
with results pointing out the stochastic nature of the algorithm. The electric
responses presented are in line with the expected results from the 
deterministic theory.
\end{abstract}

\maketitle

\section{Introduction}
\label{sec:int}

    Historically, chemical kinetics is formulated as a set of equations in
order to predict the time evolution of the concentration of a number of
species undergoing a set of reactions. With the assumption that the
concentrations $\mathbf{Z} = \mathbf{Z}(t)$ can be well-described by a set of
ordinary differential equations, the so-called Reaction Rate Equations (RRE)
became an important and useful tool in chemistry. As more complex systems
emerged, its studies demanded more mathematical tools, and several
approximations were developed, from steady-state analytical solutions to
spatial functions solved by numerical methods. Underlying it all is the
understanding that, although all chemical processes are discrete and occur in
the atomic and molecular scale, its macroscopic behavior can be reasonably
explained and reproduced with continuous functions.

    However, this macroscopic, continuous approach is ultimately just an
approximation, albeit a very good one in most usual conditions (for noteworthy
remarks on this point, see \cite{maioli16}). On a different front, it was
pursued the microscopical, statistical formulation that would reduce to the RRE
in the macroscopic limit (for a proper introduction and review on this topic,
see \cite{mcquarrie67}). In the framework of statistical mechanics, the
stochastic approach to chemical kinetics was developed, where reactions are
taken as random, discrete changes in the system, following a successful
interaction between species. This approach is marked by the postulation of the
chemical master equation (CME), a Fokker-Planck-type equation describing the
time evolution of the probability density function of finding the system on a
given configuration. Similarly to the RRE, if not worse, most systems of
interest have no analytical solution to the master equation, and numerical
methods proved inefficient for these systems.

    This was partly solved by the development of the Stochastic Simulation
Algorithm (SSA) by Gillespie\cite{gillespie76}. Originally being a framework
derived logically from statistical arguments, Gillespie went on to prove the
connection of the SSA and the CME, providing a detailed derivation of the
latter\cite{gillespie92}. The SSA provides the time evolution of a system
described by a CME, in a discrete-jump Markov-chain approximation. The result
is a time trajectory in configuration space, noteworthy for producing results
statistically equal to the RRE\cite{gillespie77}. The SSA proved particularly
useful for biological systems, where very low concentrations of enzymes and
other macromolecules interact in complex mechanisms\cite{gillespie13}. However
certain systems presented large differences in the time scale of its various
processes, significantly raising the computational cost for its simulation and
reducing the usefulness of the exact approach. As the field progressed, several
approximations and methods were developed, leading to a large array of
algorithms enabling the simulation of diverse and complex
systems\cite{gillespie01,gillespie07,gillespie09a,gillespie13}. The development
of the field led to the derivation of a chemical Langevin
equation\cite{gillespie00}, and ultimately to a rigorous connection between
both microscopic and macroscopic approaches\cite{gillespie09b}, showing that
the CME reduces to the RRE in the thermodynamic limit.

    It must be noted, however, that the SSA was originally proposed for
reactions in gas phase, and later on extended to diluted systems in a solvent,
remaining apparently underused in fields outside biochemistry. As complex,
non-linear systems are explored for a variety of objectives, and low
concentrations are used in ever more sensitive systems, the deterministic
results of the RRE fails to capture the intrinsic randomness of such systems.
This is particularly striking for fields relying on interfacial systems,
notably heterogeneous catalysis and electrochemistry. Nanometer-sized particles
and complex feedback loops lead to systems sensitive to fluctuations, and
advanced probing techniques respond accordingly, resulting in a ``noise''
mostly associated to thermal fluctuations (e.g. \cite{bard96,dick15,kim16} and
references therein). The stochastic approach is, therefore, not only
appropriate to study such systems, but one could argue that it is required in
most of them, as the SSA proved to be in biological systems. A variety of
frameworks and algorithms have been proposed to address this, however they are
usually connected to advanced molecular dynamics and electronic structure
methods, being cumbersome and unfamiliar to most experimental chemists. In this
way, the main objective of this work is to propose an extension of Gillespie's
SSA to interfacial systems, in particular electrochemical ones, providing a
straightforward framework to back up microscopic experiments. The remainder of
this work is divided as follows. In Section \ref{sec:cme} it is provided the
physical assumptions and justifications that lead to the master equation in an
interfacial system. Section \ref{sec:ssa} goes through the stochastic algorithm
and the proposed response for electrochemical systems, as well as the details
of the implementation. Section \ref{sec:res} shows some numerical results of
the current implementation for simple electrochemical systems. Finally, Section
\ref{sec:sum} summarizes the work and some perspectives are pointed out.

\section{The Chemical Master Equation}
\label{sec:cme}

    The aim of this section is to derive, heuristically, the CME that
describes the time evolution of the probability density function of the
population vector $\mathbf{X}(t)$ of a system representing the surface of
a material and the chemical species adsorbed to it. In principle, there is no
need to re-derive the master equation, since it is supposed to be able to
describe any reasonably well-behaved chemical system. The actual point of this
derivation is to provide justifications for the assumptions made, and therefore
identify the physical meaning of the constraints required for it. The train of
thought follows closely that of Gillespie's in \cite{gillespie92}.

    Consider a bi-dimensional system consisting of $N$ adsorbed chemical species
$\mathcal{S}_1, \mathcal{S}_2,\ldots,\mathcal{S}_N$, which interact through $M$
elementary reaction channels $\mathcal{R}_1,\mathcal{R}_2,\ldots,
\mathcal{R}_M$. Since it is the surface of a material, the different adsorption
sites could be identified as ``chemical'' species, or ``holes'', however the
generality on the identity of species will be maintained. The system is, in
general, open, consisting of a surface of area $A$, not necessarily flat, in
thermal equilibrium at absolute temperature $T$. The system is described at any
given time by the state vector $\mathbf{X}(t)=\mathbf{n}$, where:
\begin{eqnarray}
    X_i(t) & \equiv & \mathrm{population\ number\ of\ species}\ \mathcal{S}_i\ \mathrm{in\ the\ system}\nonumber \\
    & & \mathrm{at\ time}\ t,\ \mathrm{with}\ i\in[1,N],
\end{eqnarray}
\noindent
$X_i(t)$ is connected with the species' surface concentration by $\Gamma_i(t) =
X_i(t)/A$.

    To justify the Markov-chain approach to stochastic kinetics, two points are
proposed. Unlike in gas-phase, the concept of a well-stirred mixture is not a
necessary consequence of thermal equilibrium, as translation in the surface is
usually dependent on hopping between energy barriers. The analogous situation
can be given by a surface where the energy barrier for site hopping has the
same order of magnitude, or lower, than the thermal energy $k_BT$, where $k_B$
is Boltzmann's constant. This would allow for an almost free translation,
similarly to the scenario depicted by the linear adsorption isotherm, i.e.
Henry's law\cite{heer71}. This covers the homogeneity of the system and
reaction channels that involves collision between species. For channels
representing reactions with the surface, two approaches might be taken:
\begin{enumerate}
    \item collision between the adsorbed species and the appropriate
    ``hole'';
    \item reaction coordinate incorporates vibrational and rotational degrees
    of freedom of involved species.
\end{enumerate}
\noindent
Approach \#1 could be understood to reasonably justify the memory-less scenario
if one accepts that translation is also sufficient to thermalize a system after
an unsuccessful, reaction-less collision between two adsorbed species. On the
other hand, approach \#2 gives a different possibility for thermalization of
the system after a tunneling attempt between adsorbed species and the surface,
namely the vibrational and rotational modes of the adsorbed species. Either
way, it is assumed that translational, in this case bi-dimensional, and/or
vibrational and rotational modes are sufficient to thermalize the system,
destroying any memory effect.

    The thermalization of the system, as mentioned above, is necessary to
enforce the following conditions: i) the position of a randomly selected
molecule can be treated as a random variable of the system, being uniformly
distributed over the surface; and ii) the energy of a randomly selected
molecule can be treated as a random variable that follows Maxwell-Boltzmann
statistics. The latter statement is broader than that originally given by
Gillespie in \cite{gillespie92}, where the molecules' momentum were taken as
the random variable, hence only the kinetic energy was considered. However
if we are to include the possibility of thermalization through vibrational and
vibrational modes, solely considering the kinetic energy is insufficient to
give an accurate physical picture of the system. The end point is the same,
though, as Gillespie's, i.e. that the molecular positions and internal energies
are statistically independent of each other.

    Having defined the system and its variables, it is now needed to establish
its time evolution. It is defined the reaction probability:
\begin{eqnarray}
    \pi_\mu(t,\mathrm{d}t) & \equiv & \mathrm{probability\ that\ a\ randomly\ selected\ combination\ of\ \mathcal{R}_\mu} \nonumber \\
    & & \mathrm{reactant\ molecules\ at\ time}\ t\ \mathrm{will\ react\ accordingly\ in\ the} \nonumber \\
    & & \mathrm{next\ infinitesimal\ time\ interval}\ [t,t+\mathrm{d}t), \mathrm{with}\ \mu\in[1,M].
\end{eqnarray}
\noindent
It is shown in \cite{gillespie92} that $\pi_\mu(t,\mathrm{d}t)$ can be written
in the form
\begin{eqnarray}
    \pi_\mu(t,\mathrm{d}t) & = & c_\mu \mathrm{d}t \label{eq:pi}
\end{eqnarray}
\noindent
for most, if not all, chemical reaction channels as a good approximation. The
term $c_\mu$ is called the transition probability, or in Gillespie's original
terminology, the specific probability rate constant. This term is independent
of $t$, and reflects the particularities of each channel $\mathcal{R}_\mu$. In
Gillespie's original work, it was sufficient to demonstrate that $\pi_\mu$ can
be written as in Eq. \ref{eq:pi}. Here, it is important to elaborate it
further. Firstly, if one considers that any surface reaction can be represented
as a collision between molecules, similarly to the gas-phase, then the
arguments given in \cite{gillespie92} should be enough to justify the use of
Eq. \ref{eq:pi}, with minor modifications due to reduced dimensionality. The
same arguments are reasonably applied even in the circumstance where the
vibrational or vibrational modes are the reaction coordinate, as one could
interpret it as a single molecule reaction, driven by its own internal
mechanism. Secondly, surface reactions actually take place in an interface
between different phases, where a population-dependent difference in free
energy arises. In electrochemistry, this is usually represented by the
electrode potential, $E = E(\mathbf{X}(t))$, being either a parameter or an
observable in electrochemical experiments. In the case of $E$ as an variable,
it is dependent on the population numbers, which makes $c_\mu = c_\mu(E(t))$,
since the populations numbers are time-dependent. This is against the
assumptions required to derive the master equation. The following reasoning is
proposed to mend this apparent incompatibility. Following basic electrochemical
theory\cite{bard01}, one starts by redefining the transition probability:
\begin{eqnarray}
    c_\mu\left(E(t)\right) & = & c_\mu^0 f\left(E(t)\right) \label{eq:f_e}
\end{eqnarray}
\noindent
where $c_\mu^0$ is now the time-independent transition probability, and
$f(E(t))$ is a function that gives the appropriate dependency on $E$. This
alone is not enough to mend the gap. To relax the dependency of $E$
with $t$, recall that the time dependency arises due to the population numbers
$\mathbf{X}(t)$. However, $\mathbf{X}$ is actually a constant between
reactions, i.e. whithin the interval $[t,t+\mathrm{d}t)$, since its time
evolution is given by discrete Markov jumps. Therefore, it is possible to
establish that
\begin{eqnarray}
    E\left(\mathbf{X}(t)\right) & = & \mathrm{constant,\ for}\ t\in[t,t+\mathrm{d}t),
\end{eqnarray}
\noindent
thus removing the dependency of $c_\mu$ on $t$ between reactions. The
actual form of $f(E)$, usually an exponential of $E$, is not actually relevant,
and can be modified according to the desired electrochemical kinetic model.
However, it is important that $E$ must either be a constant parameter or
observable of the system, i.e. only potentiostatic (constant applied electrode
potential) or galvanostatic (constant applied electric current) setups are
relevant. To allow for time-dependent $E$ as a parameter, it is necessary to
resort to approximations.

    Now, it should be noted that Eq. \ref{eq:pi} defines the probability that
a randomly selected combination of molecules react through channel
$\mathcal{R}_\mu$. To calculate the actual probability of observing a reaction
through channel $\mathcal{R}_\mu$, it is necessary to account for all possible
combinations of molecules that might take part in it. This is done by defining
the function
\begin{eqnarray}
    h_\mu\left(n_1,\ldots,n_N\right) & \equiv & \mathrm{the\ number\ of\ distinct\ combinations\ of}\ \mathcal{R}_\mu\ \mathrm{reactant}\nonumber \\
  & & \mathrm{molecules\ in\ the\ system\ when\ there\ are\ exactly}\nonumber\\
  & & n_i\ \mathrm{molecules\ of\ species}\ \mathcal{S}_i,\ \mathrm{with}\ i\in[1,N]
\end{eqnarray}
\noindent
The actual form of $h_\mu$ is given by the state change matrix
$\boldsymbol{\nu}$, defined as
\begin{eqnarray}
    \nu_{\mu i} & \equiv & \mathrm{the\ change\ in}\ X_i\ \mathrm{caused\ by\ the\ occurrence}\nonumber\\
    & & \mathrm{of\ one}\ \mathcal{R}_\mu\ \mathrm{reaction.}
\end{eqnarray}
\noindent
At this point we diverge somewhat from \cite{gillespie92} by further describing
it as:
\begin{eqnarray}
    \boldsymbol{\nu} & = & \mathbf{P} - \mathbf{R}
\end{eqnarray}
\noindent
where $\mathbf{P}$ and $\mathbf{R}$ are the products and reactants
stoichiometric coefficients matrices, respectively. This implicitly defines
all stoichiometric coefficients as positive integers, with the correct change in
$\mathbf{X}$ given by the above definition of $\boldsymbol{\nu}$. This is done
in order to provide a general definition for $h_\mu$. By inspecting the form
of elementary reaction steps, it is possible to generalize and write:
\begin{eqnarray}
    h_\mu(\mathbf{n}) & = & \prod_{i=1}^N\binom{n_i}{R_{\mu i}} \nonumber \\
          & = & \prod_{i=1}^N\frac{n_i!}{R_{\mu i}!\left(n_i-R_{\mu i}\right)!} \label{eq:h}
\end{eqnarray}
\noindent
where it becomes clear why it is necessary that $\mathbf{R}$ consists only of
positive integers. Finally, then, one finds that the total probability of a
given combination of molecules reacting through channel $\mathcal{R}_\mu$ in
the time interval $[t,t+\mathrm{d}t)$ is given by the term
$c_\mu(E)h_\mu(\mathbf{n})\mathrm{d}t$.

    Having gone through the same development as Gillespie did in
\cite{gillespie92}, one can use the same three theorems in order to conclude
this derivation:
\begin{enumerate}
    \item the probability of one reaction $\mathcal{R}_\mu$ to occur in the
          system in the time interval $[t,t+\mathrm{d}t)$, given that
          $\mathbf{X}(t) = \mathbf{n}$, is $c_\mu(E)h_\mu(\mathbf{n})
          \mathrm{d}t + O\left(\mathrm{d}t^2\right)$;
    \item the probability of no reaction occuring in the interval $[t,t+
          \mathrm{d}t)$, given $\mathbf{X}(t)=\mathbf{n}$, is $1-\sum_{\mu=1}
          ^M c_\mu(E)h_\mu(\mathbf{n})\mathrm{d}t + O\left(\mathrm{d}t^2
          \right)$;
    \item the probability of an $m$ number of reactions occurring in the system
          in the time interval $[t,t+\mathrm{d}t)$ is $O\left(\mathrm{d}t^m
          \right)$.
\end{enumerate}
\noindent
The only remarks that might be reinforced is that trying to determine the time
evolution of the species' population vector $\mathbf{X}(t)$ is hopeless in most
cases. Instead, one defines
\begin{eqnarray}
    \mathcal{P}\left(\mathbf{n},t|\mathbf{n}_0,t_0\right) & = & \mathrm{probability\ that}\ \mathbf{X}(t) = \mathbf{n},\ \mathrm{given\ that}\ \mathbf{X}(t_0) = \mathbf{n}_0\nonumber\\
    & & \mathrm{and}\ t \geq t_0,
\end{eqnarray}
\noindent
and try to determine its time evolution. Using the theorems above to account
for the probabilities that the species' population numbers be $\mathbf{X}(t) =
\mathbf{n}$, during a time interval $[t,t+\mathrm{d}t)$, one obtains:
\begin{eqnarray}
    \mathcal{P}\left(\mathbf{n},t+\mathrm{d}t|\mathbf{n}_0,t_0\right) & = & \mathcal{P}\left(\mathbf{n},t|\mathbf{n}_0,t_0\right)\left(1 - \sum_{\mu = 1}^M c_\mu(E)h_\mu(\mathbf{n})\mathrm{d}t + O(\mathrm{d}t^2)\right) \nonumber \\
    & & + \sum_{\mu =1}^M \mathcal{P}\left(\mathbf{n}-\boldsymbol{\nu}_\mu,t|\mathbf{n}_0,t_0\right)\left[c_\mu(E)h_\mu(\mathbf{n}-\boldsymbol{\nu}_\mu)\mathrm{d}t + O(\mathrm{d}t^2)\right] \nonumber \\
    & & + O(\mathrm{d}t^2)
\end{eqnarray}
\noindent
where $\boldsymbol{\nu}_\mu$ is the state change vector for channel $\mathcal{R}
_\mu$. This equation indicates that, at first order in $\mathrm{d}t$, two
possibilities exist for the system to reach state $\mathbf{X}(t+\mathrm{d}t)=
\mathbf{n}$: i) the system is at state $\mathbf{X}(t)=\mathbf{n}$ and no
reaction occurs, and ii) the system is at state $\mathbf{X}(t)=\mathbf{n}-
\boldsymbol{\nu}_\mu$ and one reaction $\mathcal{R}_\mu$ occurs. Finally, one
subtracts the probability at time $t$, divides by $\mathrm{d}t$, and taking
the limit $\mathrm{d}t\rightarrow 0$, one obtains the following equation:
\begin{eqnarray}
    \frac{\partial}{\partial t}\mathcal{P}\left(\mathbf{n},t|\mathbf{n}_0,t_0\right) & = & \sum_{\mu =1}^M\left[c_\mu(E)h_\mu(\mathbf{n}-\boldsymbol{\nu}_\mu)\mathcal{P}(\mathbf{n}-\boldsymbol{\nu}_\mu,t|\mathbf{n}_0,t_0)\right. \nonumber\\
    & &  \left. - c_\mu(E)h_\mu(\mathbf{n})\mathcal{P}(\mathbf{n},t|\mathbf{n_0},t_0)\right]
\end{eqnarray}
\noindent
subjected to the initial conditions
\begin{eqnarray}
    \mathcal{P}\left(\mathbf{n},t=t_0|\mathbf{n}_0,t_0\right) & = & \begin{cases} 1, & \mathrm{if}\ \mathbf{n} = \mathbf{n}_0 \\ 0, & \mathrm{if}\ \mathbf{n} \neq \mathbf{n}_0 \end{cases}
\end{eqnarray}
This is the so-called chemical master equation, as derived by Gillespie in
\cite{gillespie92}. It follows then, by the arguments given by him, that the
Stochastic Simulation Algorithm (SSA)\cite{gillespie76} can also be applied to
surface reactions, given that the above constraints are observed, particularly
the low energy barrier for site hopping and that the electrode potential be
constant between reactions.

\section{The Stochastic Simulation Algorithm}
\label{sec:ssa}

\subsection{Mathematical Framework}
\label{ssec:math}

    The SSA is deemed as a logic equivalent to the CME\cite{gillespie92},
differing by the fact that, while the CME describes the time evolution of the
probability density function, the SSA provides a time trajectory in population
space. This is accomplished by sampling the joint probability distribution:
\begin{eqnarray}
    \mathcal{P}\left(\tau,\mu|\mathbf{n},t\right)\mathrm{d}\tau & \equiv & \mathrm{probability\ that\ the\ next\ reaction\ in\ the\ system\ will\ occur} \nonumber \\
    & & \mathrm{in\ the\ time\ interval}\ [t+\tau,t+\tau+\mathrm{d}\tau)\ \mathrm{and\ will\ be\ an} \nonumber \\
    & & \mathcal{R}_\mu\ \mathrm{reaction.}
\end{eqnarray}
\noindent
To derivate a formula for this joint probability, it is defined a given
channel's reaction propensity and the total reaction propensity functions:
\begin{eqnarray}
    a_\mu\left(\mathbf{n},E\right) & = & c_\mu\left(E\right)h_\mu\left(\mathbf{n}\right) \label{eq:a_mu}\\
    a_0 \left(\mathbf{n},E\right) & = & \sum_{\mu =1}^M a_\mu\left(\mathbf{n},E\right) \label{eq:a0}
\end{eqnarray}
\noindent
respectively, both with units of $\mathrm{s^{-1}}$. These functions are
connected to Theorems \#1 and \#2: $a_\mu$ representing the probability per
unit time, to first order in $\mathrm{d}t$, that channel $\mathcal{R}_\mu$ will
fire in the time interval $[t,t+\mathrm{d}t)$; while the probability per unit
time, also to first order in $\mathrm{d}t$, of no reaction occurring in the
time interval $[t,t+\mathrm{d}t)$ is given by $1-a_0$.

    Considering this, it can be shown that the joint probability can be written
as\cite{gillespie92}
\begin{eqnarray}
    \mathcal{P}\left(\tau,\mu|\mathbf{n},t\right) & = & a_\mu\left(\mathbf{n},E\right)\mathrm{exp}\left[-a_0\left(\mathbf{n},E\right)\tau\right]
\end{eqnarray}
\noindent
In this way, the SSA proceeds by sampling the time for the next reaction,
$\tau$, and the channel that will subsequently fire, $\mu$. This is done by
noting that the time for \emph{any} reaction to occur in the system is given
by:
\begin{eqnarray}
    \mathcal{P}\left(\tau|\mathbf{n},t\right) & = & \sum_{\mu =1}^M \mathcal{P}\left(\tau,\mu|\mathbf{n},n\right) \nonumber \\
    & = & a_0\left(\mathbf{n},E\right)\mathrm{exp}\left[-a_0\left(\mathbf{n},E\right)\tau\right]
\end{eqnarray}
\noindent
and the probability of a given channel $\mu$ firing is simply given by
$a_\mu/a_0$. Details about the recommended implementation are described in
\cite{gillespie76}.

    As noted by the explicit dependency of the reaction propensity function
in Eq. \ref{eq:f_e}, the SSA works exactly the same way for surface reactions,
electrochemical or not. This was noted in the beginning of the CME derivation.
The interest now lies in providing the associated electric response for the
electrochemical reactions under the relevant conditions:
\begin{itemize}
    \item Potentiostatic, with constant applied electrode potential $E$, the
    electric current $I=I\left(t;E\right)$.

    \item Galvanostatic, with constant applied electric current $I$, the
    electrode potential $E=E\left(\mathbf{X}(t);I\right)$.
\end{itemize}

    For the potentiostatic case, the most obvious solution would be to simply
account for the number of electrons transferred in unit time at a given
electrode potential, and then provide the electric current:
\begin{eqnarray}
    I & = & \frac{\delta Q}{\delta t}
\end{eqnarray}
\noindent
where $Q$ is the total charge transferred between the system's population and
the surface. However, in this way, there's no clear way to simulate
galvanostatic systems. To circumvent this, a different interpretation is
proposed. For an ensemble, Eqs. \ref{eq:a_mu} and \ref{eq:a0} can be understood
as the average number of reactions per unit time for channel $\mathcal{R}_\mu$
and all channels, respectively. By considering that electrochemical channels
involve electron transfer to or from the electrode, it is straightforward, to
connect these to the electric current for the ensemble:
\begin{eqnarray}
    I_\mu\left(\mathbf{n},E\right) & = & q_e\nu^e_\mu a\left(\mathbf{n},E\right) \\
    I\left(\mathbf{n},E\right) & = & \sum_{\mu = 1}^M I_\mu \nonumber \\
    & = & q_e\sum_{\mu =1}^M \nu^e_\mu a_\mu\left(\mathbf{n},E\right) \label{eq:I}
\end{eqnarray}
\noindent
where $q_e$ is the fundamental charge, $\nu^e_\mu=\pm 1$ is the number of
electrons transferred in channel $\mathcal{R}_\mu$. In this way, it is defined
$\nu^e_\mu = 1$ for oxidation and $\nu^e_\mu = -1$ for reduction reactions, and
therefore $I_\mu > 0$ and $I_\mu < 0$ for anodic and cathodic currents,
respectively. Equation \ref{eq:I} directly provides the desired electric
response in the case of potentiostatic setups, while providing the electrode
potential through Eq. \ref{eq:f_e}, in the case of galvanostatic setups. It is
noteworthy that no electrochemical kinetic model has been assumed so far, so
the form of $f(E)$ is completely arbitrary, as long as it follows the required
thermodynamical constraints.

    The above ensemble can be shown to be consistent with the thermodynamic
limit of the CME/SSA approach. It was shown in \cite{gillespie00} that, through
the conditions used in the $\tau$-leap method\cite{gillespie01}, it is possible
to describe the time evolution of the population vector $\mathbf{X}(t)$ through
a Langevin-type equation. The $\tau$-leaping conditions are:
\begin{eqnarray}
    a_\mu(\mathbf{n},E) & \approx & \mathrm{constant\ in}\ [t,t+\tau),\ \forall \mu \label{eq:tau1} \\
    a_\mu(\mathbf{n},E)\tau & \gg & 1,\ \forall \mu \label{eq:tau2}
\end{eqnarray}
\noindent
Assuming that both conditions be satisfied, the time evolution of the
population vector can be written as\cite{gillespie13}:
\begin{eqnarray}
    \mathbf{X}(t+\mathrm{d}t) - \mathbf{X}(t) & = & \sum_{\mu =1}^M \boldsymbol{\nu}_\mu a_\mu\left(\mathbf{X}(t),E\right)\mathrm{d}t \nonumber \\
     & & + \sum_{\mu =1}^{M}\boldsymbol{\nu}_\mu \mathcal{N}_\mu(0,1)\sqrt{a_\mu\left(\mathbf{X}(t),E\right)\mathrm{d}t} \label{eq:cle}
\end{eqnarray}
\noindent
where $\mathcal{N}(0,1)$ is a normally distributed random number with zero mean
and unity standard deviation. Equation \ref{eq:cle} is the chemical Langevin
equation, as derived by Gillespie. It has been shown that the $\tau$-leap
conditions can always be satisfied\cite{gillespie09b}, given that the system be
made ``sufficiently large''\cite{gillespie13}, therefore Eq. \ref{eq:cle} is
always valid. Taking the thermodynamic limit, that is, increasing the system
such that $A \rightarrow \infty$ with the constraint that the surface
concentrations $\boldsymbol{\Gamma} = \mathbf{X}/A$ remains constant; the term
on the left side of Eq. \ref{eq:cle} grows linearly with system size, and so
does the first term on the right side. The second term, the ``noise'', grows
with the square root of the size, being negligible in the thermodynamic limit.
Hence, Eq. \ref{eq:cle} reduces to
\begin{eqnarray}
    \frac{\mathrm{d}\mathbf{X}(t)}{\mathrm{d}t} & = & \sum_{\mu =1}^{M}\boldsymbol{\nu}_\mu a_\mu\left(\mathbf{X}(t),E\right) \label{eq:rre}
\end{eqnarray}
\noindent
which is the usual, deterministic approach to chemical kinetics, the Reaction
Rate Equation (RRE), more usually written using surface concentrations when
dealing with electrochemical reactions:
\begin{eqnarray}
    \frac{\mathrm{d} \boldsymbol{\Gamma}(t)}{\mathrm{d}t} & = & \frac{\mathrm{d}}{\mathrm{d}t}\left(\frac{\mathbf{X}(t)}{A}\right) \nonumber\\
    & = & \sum^M_{\mu =1}\boldsymbol{\nu}_\mu a'_\mu\left(\boldsymbol{\Gamma}(t),E\right)
\end{eqnarray}
\noindent
where $a'_\mu = a_\mu/A$. Considering the possibility of electron transfer to
or from the surface, one writes Faraday's law of electrolysis\cite{bard01}:
\begin{eqnarray}
    I & = & q_e A\frac{\mathrm{d}\boldsymbol{\Gamma}(t)}{\mathrm{d}t}\nonumber\\
    & = & q_e\sum_{\mu =1}^{M}\nu^e_\mu a_\mu\left(\mathbf{X}(t),E\right)
\end{eqnarray}
\noindent
which is the same as Eq. \ref{eq:I}, thereby justifying the ensemble
interpretation proposed above.

\subsection{Implementation}
\label{ssec:imp}

\begin{figure}[h]
    \begin{center}
        \includegraphics[width=0.8\textwidth]{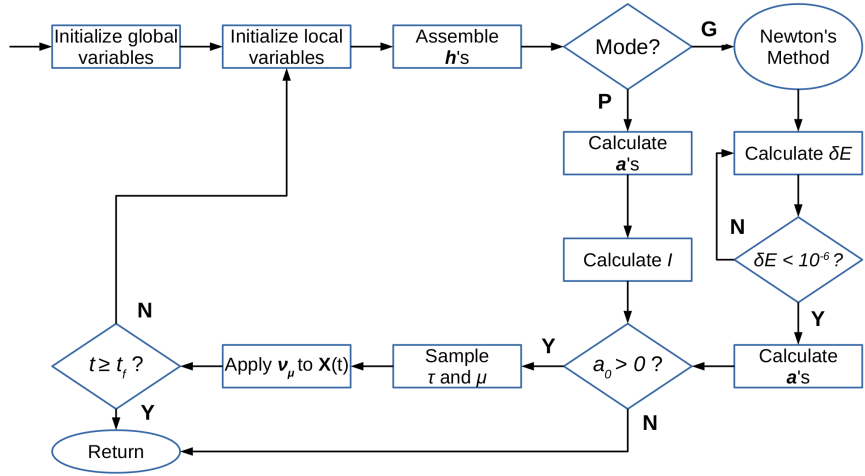}
        \caption{
            Flowchart of the current implementation of the electrochemical
            SSA.
        }
        \label{fig:1}
    \end{center}
\end{figure}

    For the results shown in Section \ref{sec:res}, a MATLAB$^\circledR$ 2012a
script was written\footnote{The MATLAB script and a Python3 version are
available in https://github.com/beruski/ec-ssa.}. As an example, the
Butler-Volmer model of electrochemical kinetic is used to define $f(E)$. The
macroscopic derivation for the model can be found in \cite{bard01}. In the
present framework, it is implemented as:
\begin{eqnarray}
    f(E) = \mathrm{exp}\left[\alpha_\mu\frac{\nu^e_\mu q_e}{k_B T}\left(E-E_\mu^{0'}\right)\right]
\end{eqnarray}
\noindent
where $\alpha_\mu$ is the transference coefficient and $E_\mu^{0'}$ is the
formal potential of channel $\mathcal{R}_\mu$. The usually employed Faraday's
constant, $F$, and the universal gas constant, $R$, have been exchanged for
its molecular counterparts, $q_e$ and $k_B$ respectively, in order to maintain
coherence. Hence, the reaction propensity function is completely defined as
\begin{eqnarray}
    a_\mu\left(\mathbf{n},E\right) & = & c_\mu^0h_\mu\left(\mathbf{n}\right)\mathrm{exp}\left[\alpha_\mu\frac{\nu^e_\mu q_e}{k_BT}\left(E-E_\mu^{0'}\right)\right] \label{eq:bv}
\end{eqnarray}
\noindent
and consequently, the associated individual and total currents. For
potentiostatic systems, the implementation is quite straightforward, with the
current in Eq. \ref{eq:I} possessing a parametric dependence on $E$:
\begin{eqnarray}
    I\left(\mathbf{n};E\right) & = & q_e\sum_{\mu =1}^M \nu^e_\mu a_\mu\left(\mathbf{n};E\right) \label{eq:pot}
\end{eqnarray}
\noindent
For galvanostatic systems, a simple minimization of Eq. \ref{eq:I}, coupled to
Eq. \ref{eq:bv}, is sufficient:
\begin{eqnarray}
    \sum_{\mu =1}^M \nu^e_\mu a_\mu\left(\mathbf{n},E\right) - \frac{I}{q_e} & = & 0 \label{eq:gal}
\end{eqnarray}
\noindent
In the current implementation, the minimization is carried through the
Newton-Raphson method, with a relative tolerance of $10^{-6}$. Another point
worth noticing is that function $h_\mu(\mathbf{n})$ is not implemented using
the definition given in Eq. \ref{eq:h}, but using individual forms to reactions
up to order 3 for a given species. Figure \ref{fig:1} shows the core
of the implementation.

\section{Numerical Results}
\label{sec:res}

\begin{figure}
    \begin{center}
        \begin{subfigure}{0.45\textwidth}
            \includegraphics[width=\textwidth]{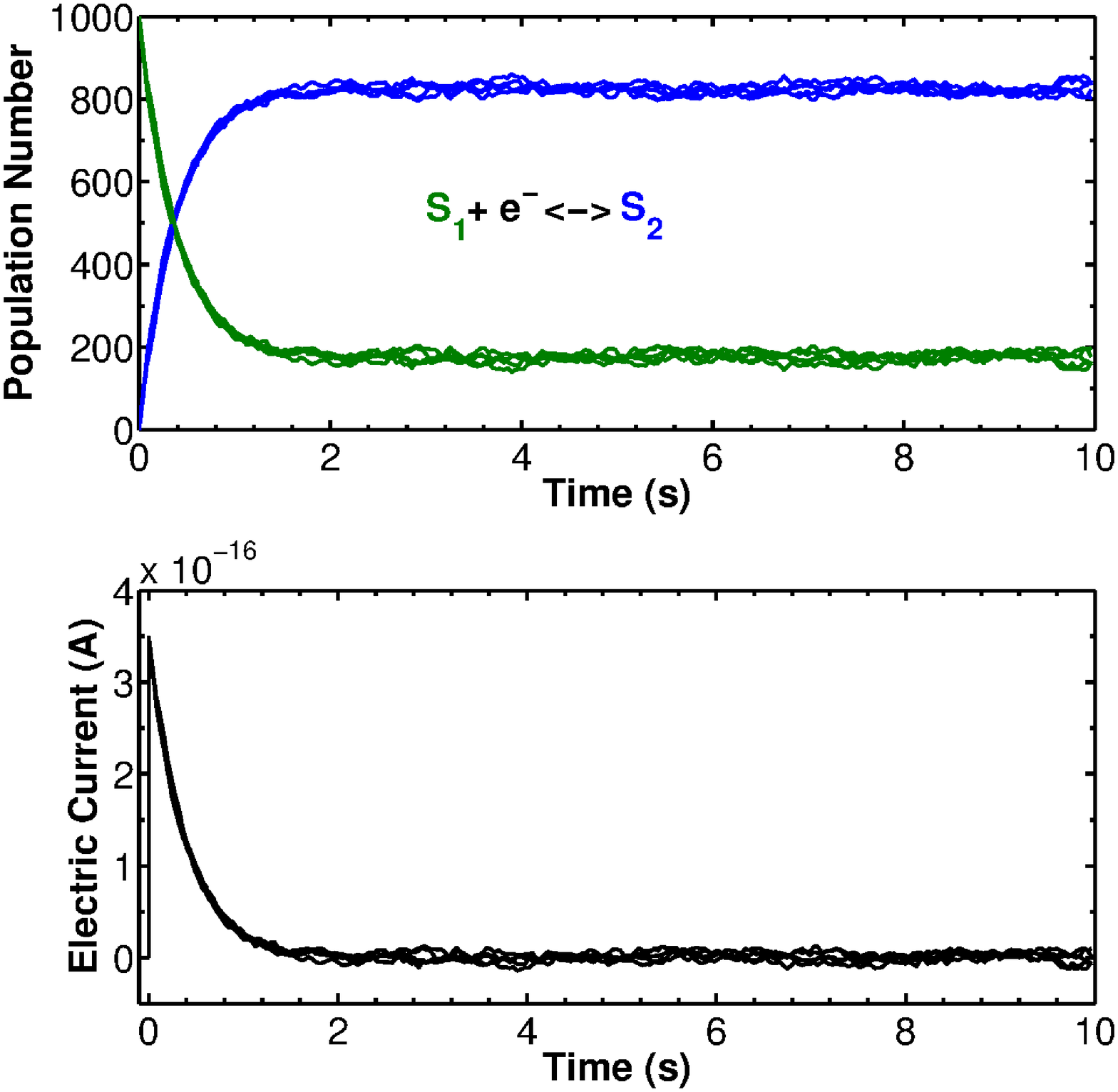}
            \caption{}
            \label{fig:2a}
        \end{subfigure}
        ~
        \begin{subfigure}{0.45\textwidth}
            \includegraphics[width=\textwidth]{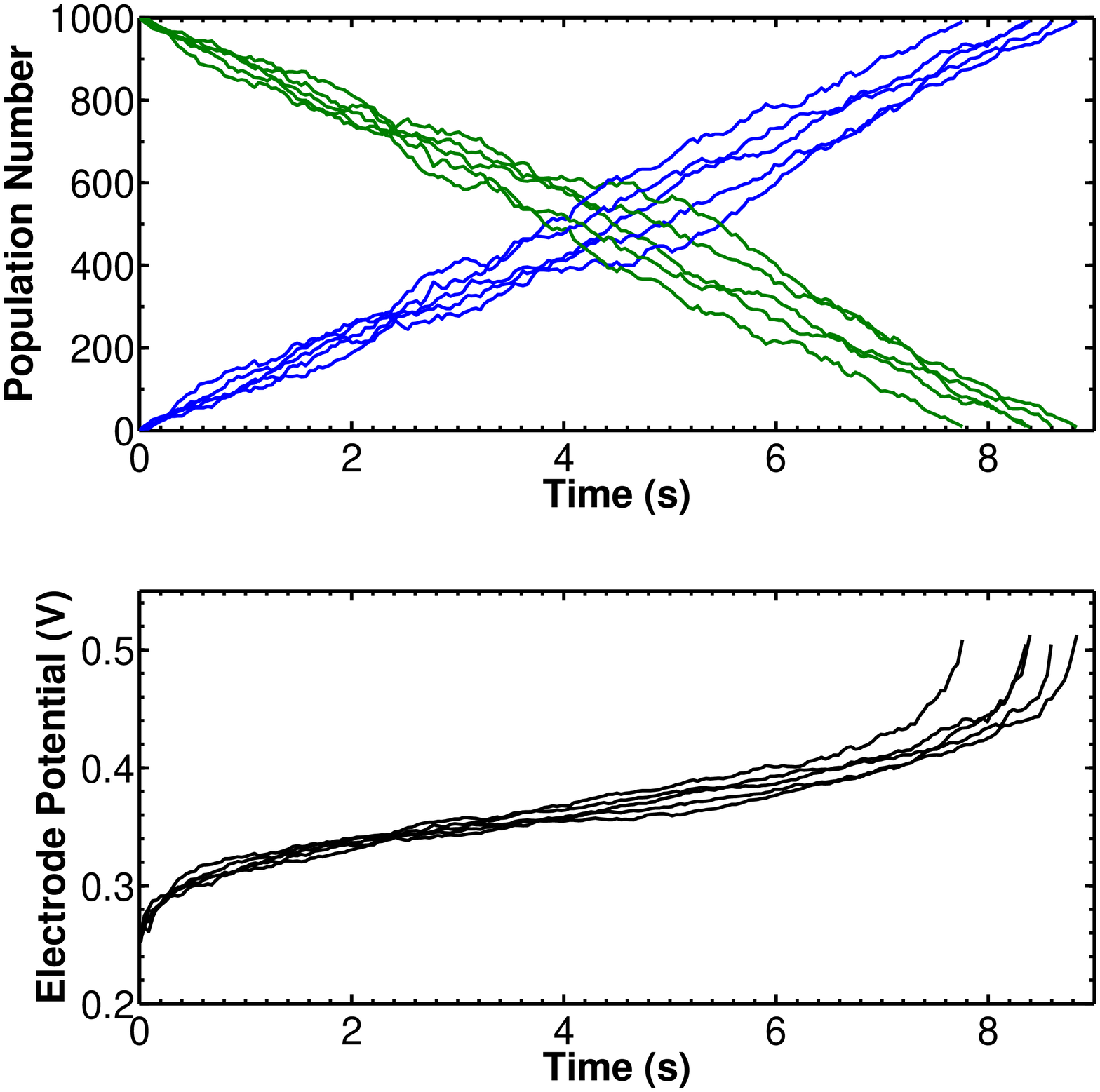}
            \caption{}
            \label{fig:2b}
        \end{subfigure}
        \caption{
            Time evolution of the system described by Eq. \ref{eq:eg1}:
            \textbf{a)} potentiostatic setup, $E=0.4\ \mathrm{V}$;
            \textbf{b)} galvanostatic setup, $I=2\times 10^{-17}\ \mathrm{A}$.
            Parameters used for both setups: $c_f^0 = c_b^0 = 1\
            \mathrm{s^{-1}}$, $\alpha_f = \alpha_b = 0.5$, $E_f^{0'}=E_b^{0'}=
            0.36\ \mathrm{V}$ and $\mathbf{X}(t=0)=[0,1000]$. Each figure shows
            $5$ superimposed independent runs.
        }
        \label{fig:2}
    \end{center}
\end{figure}

    As illustration of the framework and the current implementation, first
consider the simple redox pair:
\begin{eqnarray}
    S_1 + e^- & \rightleftharpoons & S_2 \nonumber
\end{eqnarray}
\noindent
which is implemented as two different reactions:
\begin{subequations}
    \label{eq:eg1}
    \begin{align}
        S_1 + e^- \rightarrow  & \ S_2 \\
        S_2 \rightarrow & \ S_1 + e^-
    \end{align}
\end{subequations}
\noindent
being the forward and backward reactions, respectively. Figure \ref{fig:2}
presents a few superimposed time evolutions, and associated electric responses,
for both potentiostatic and galvanostatic systems. The parameters used are
$c_f^0 = c_b^0 = 1\ \mathrm{s^{-1}}$, for the forward (reduction) and backward
(oxidation) reactions, respectively; $\alpha_f = \alpha_b = 0.5$ and $E_f^{0'}=
E_b^{0'} = 0.36\ \mathrm{V}$. The electrode potential used for Fig.
\ref{fig:2a} is $E = 0.4\ \mathrm{V}$, while the electric current used for Fig.
\ref{fig:2b} is $I = 2\times 10^{-17}\ \mathrm{A}$. The initial conditions for
both systems is $X_1(t=0) = 0$ for species $\mathcal{S}_1$ and $X_2(t=0)=1000$
for $\mathcal{S}_2$.

    It can be seen from Fig. \ref{fig:2a} that the potentiostatic
setup results in a time evolution quite similar to that of a simple chemical
equilibrium. That is expected from the effect of the electrode potential as
given in Eq. \ref{eq:bv}. The electric response is also what is expected from
the current decay of a potential-step chronoamperometry\cite{bard01}, i.e. a
high current surge followed by a fast decrease to zero at equilibrium. These
results are mainly the result of the electrochemical kinetics model chosen,
which is known to correctly reproduce most experimental results\cite{bard01}.
The main feature here would be the current oscillations around zero, which are
the result of the dynamical, and stochastic, nature of chemical equilibrium.

    On the other hand, the galvanostatic setup (Fig. \ref{fig:2b}) shows
a linear time evolution for the population numbers. Such result is relatively
unfamiliar even in electrochemistry textbooks (e.g. \cite{bard01} and
\cite{bockris02}), but it can be predicted for simple systems, such as Eq.
\ref{eq:eg1}, using Eq. \ref{eq:gal}. The electrode potential, on the other
hand, is the textbook example for a current-step galvanostatic transient,
loosely called a chronopotentiometry, showing the inflection point
approximately at $0.36\ \mathrm{V}$ and $X_1 \approx X_2$. The fluctuations are
more evident than in the potentiostatic case, although being mostly due to the
forced condition of the galvanostatic setup. This is reflected as large
differences in the electrode potential, due to the exponential nature of the
Butler-Volmer model.

\begin{figure}
    \begin{center}
        \includegraphics[width=0.5\textwidth]{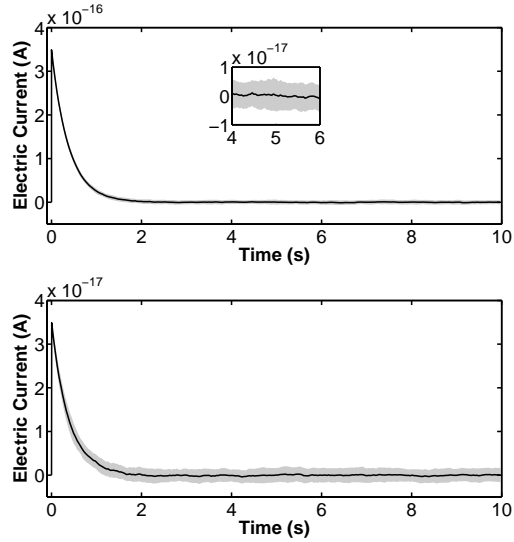}
        \caption{
            Average current associated with Eq. \ref{eq:eg1}, with one standard
            deviation shown as the shaded area: \textbf{Top:} $\mathbf{X}(t=0)=
            [0,1000]$, with inset showing a zoom in the $y$ axis;
            \textbf{Bottom:} $\mathbf{X}(t=0)=[0,100]$. Statistics for each
            figure performed over 100 runs.
        }
        \label{fig:3}
    \end{center}
\end{figure}

    A major advantage of the SSA is the possibility of obtaining statistics on
the system under study, for instance average values and expected standard
deviation. This allows a rigorous comparison with experimental results, but
also provides a picture on the effects of random fluctuations in the system.
Figure \ref{fig:3} shows the average and standard deviation of the
electric current for 100 runs of the potentiostatic setup for Eq. \ref{eq:eg1},
for two different initial population for species $\mathcal{S}_2$. In this way,
it is possible to establish if differences between a model and the experimental
results are significant, given a certain statistical power. This becomes
particularly important for ``small'' systems, as can be seen comparing the top
and bottom graphs in Fig. \ref{fig:3}. The currents differ by the
size of the system, by one order of magnitude, as is noticed by the scales of
the current. As expected, the behavior of the average current is basically the
same for both systems. However, the standard deviation associated with the
smaller system is significantly higher, relative to its peak current values,
being roughly the same as for the larger system ($\delta I\approx 0.5\times
10^{-17}\ \mathrm{A}$).

\begin{figure}
    \begin{center}
        \begin{subfigure}{0.45\textwidth}
            \includegraphics[width=\textwidth]{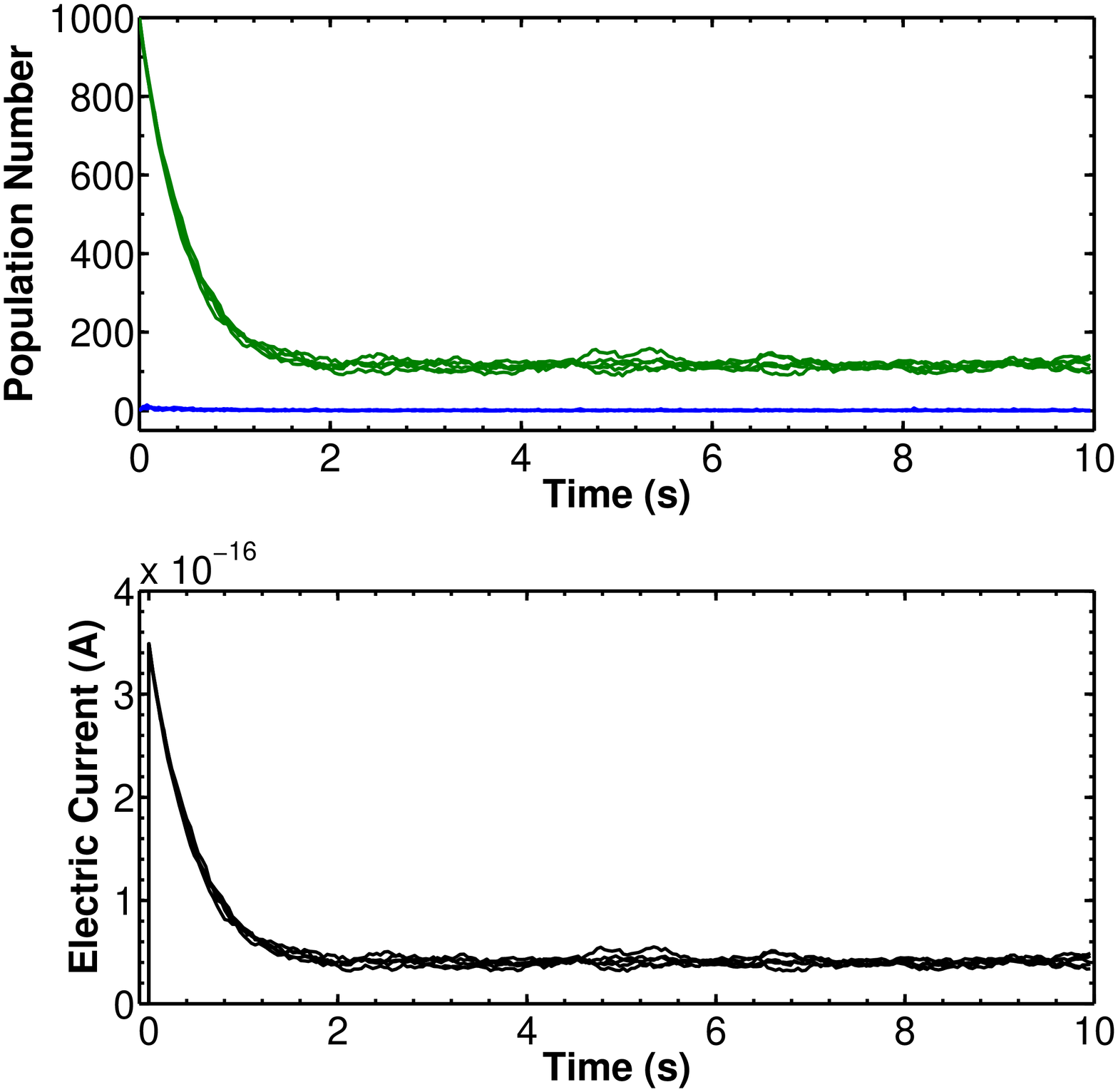}
            \caption{}
            \label{fig:4a}
        \end{subfigure}
        ~
        \begin{subfigure}{0.45\textwidth}
            \includegraphics[width=\textwidth]{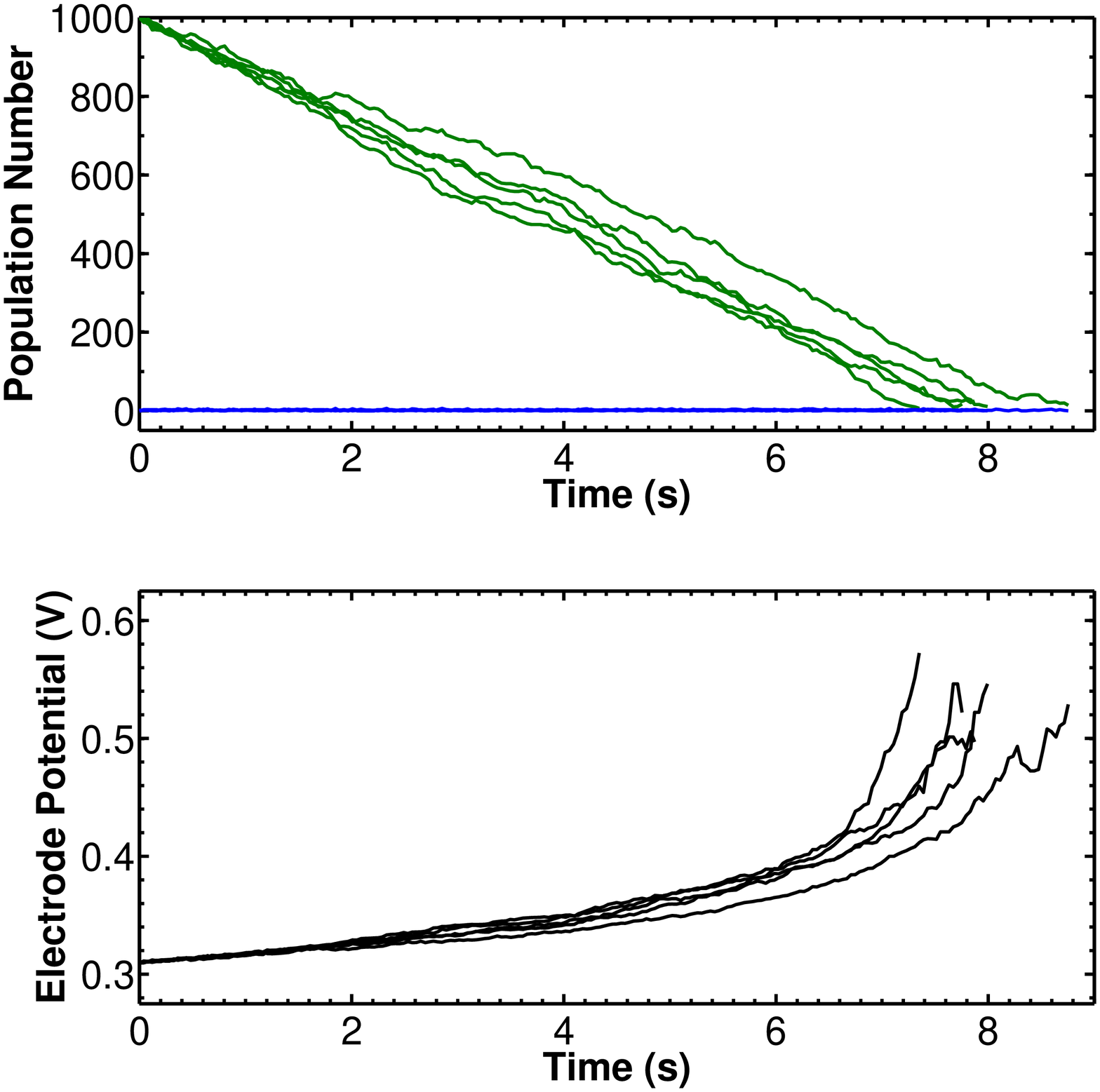}
            \caption{}
            \label{fig:4b}
        \end{subfigure}
        \caption{
            Time evolution of the system described by Eq. \ref{eq:eg2}:
            \textbf{a)} potentiostatic setup, $E=0.4\ \mathrm{V}$;
            \textbf{b)} galvanostatic setup, $I=6\times 10^{-17}\ \mathrm{A}$.
            Parameters used for both setups: $c_{b,f}^0 = c_{b,b}^0 = 1\
            \mathrm{s^{-1}}$, $\alpha_f = \alpha_b = 0.5$, $E_f^{0'}=E_b^{0'}=
            0.36\ \mathrm{V}$, $c_a^0 = c^0_c = 250\ \mathrm{s^{-1}}$
            and $\mathbf{X}(t=0)=[0,1000]$. The figures shows $5$ superimposed
            independent runs.
        }
        \label{fig:4}
    \end{center}
\end{figure}

    The simple model represented by Eq. \ref{eq:eg1} do not exhibit many of the
effects commonly seen in electrochemical systems, most notably diffusion. Being
a stochastic process, diffusion is readily incorporated in the current
framework. However, the differences in time scales that arise lead to
oversampling of the diffusion, greatly increasing the computation times for
similar sampling of the reaction events when neglecting diffusion. A
diffusional propensity function, although approximate, has been
developed\cite{gillespie09a}. Its implementation, however, requires further
development of the present framework, which are under pursue. On the other
hand, the following model shows similar effects:
\begin{subequations}
    \label{eq:eg2}
    \begin{align}
        \emptyset \rightarrow & \ S_2 \\
        S_2 \rightleftharpoons & \ S_1 + e^- \\
        S_1 \rightarrow & \ \emptyset
    \end{align}
\end{subequations}
\noindent
where the empty set means a connection with a species reservoir, with no
interest to the simulation. The redox reaction is the same as Eq. \ref{eq:eg1},
and the parameters used are the same. For the diffusion channels, only
the potential-independent specific transition coefficients are needed, being
set up as $c^0_a = c^0_c = 250\ \mathrm{s^{-1}}$. Figure \ref{fig:4} presents
the time evolution for both potentiostatic and galvanostatic setups.

    The noteworthy effect of including the diffusion channels in the system is
achieving a non-zero stationary current, in the potenstiostatic setup (Fig.
\ref{fig:4a}). This is expected from the physical scenario and the
deterministic theory\cite{bard01}. However, the decay in current does not
follows the prediction of Cottrell's equation\cite{bard01}, which predicts a
decay with $t^{-1/2}$, while the observed behavior is roughly proportional to
$t^{-1}$. This is probably due to the fact that actual diffusion depends on the
gradient of species' population, unlike what has been used in this system. For
the galvanostatic setup (Fig. \ref{fig:4b}) no significant changes are seen,
only the increased lifetime of species $\mathcal{S}_2$, as noted by the higher
value of applied current used. It is expected that more complex systems do
require diffusion in order to accurately simulate them, demanding
approximations in order to be able to perform it efficiently.

\begin{figure}
    \begin{center}
        \includegraphics[width=0.5\textwidth]{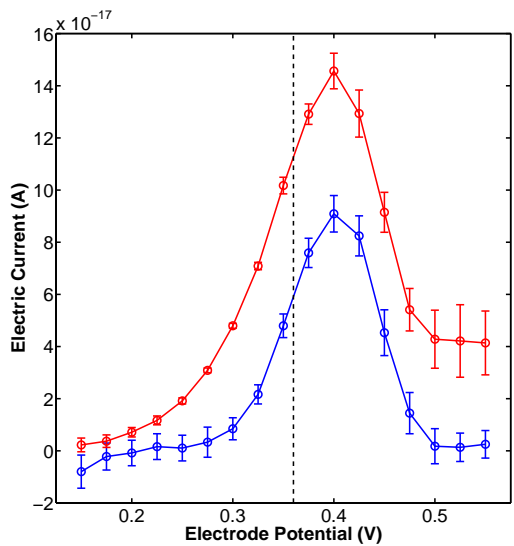}
        \caption{
            Current-sampled voltammograms with sampling time $t \approx 0.5\
            \mathrm{s^{-1}}$, for the systems of Eqs. \ref{eq:eg1}
            (\textcolor{blue}{$\circ$}) and \ref{eq:eg2}
            (\textcolor{red}{$\Box$}). Parameters are the same as described in
            Figs. \ref{fig:2} and \ref{fig:4} for Eqs. \ref{eq:eg1} and
            \ref{eq:eg2}, respectively. Statistics performed
            over 25 samples at each potential step, with error bars
            representing one standard deviation. Black dashed line shows the
            formal equilibrium potential $E^{0'}=0.36\ \mathrm{V}$.
        }
        \label{fig:5}
    \end{center}
\end{figure}

    Despite the limitations on constant applied electrode potential or electric
current, more elaborated setups can be achieved by combining different applied
parameters and sampling the associated response. The potentiostatic setup, for
instance, can be used to simulate a more commonly used experimental approach,
which consists of sampling the current at a given time, for several values of
the applied electrode potential. Such current-sampled voltammogram, as it is
called, is presented in Figure \ref{fig:5} for both Eqs. \ref{eq:eg1} and
\ref{eq:eg2}. The results presentation differs by showing a graph of $I$ as a
function of the applied electrode potential $E$, resulting in a richer source
of qualitative and quantitative information. The simulated results follows the
deterministic response\cite{bard01}: a peak current value associated to $E$
values higher than the formal potential $E^{0'}$; and a non-zero current after
the peak, for the system of Eq. \ref{eq:eg2}. This so-called diffusional
plateau is a noteworthy feature of diffusion-limited electrochemical processes,
and it is seen to be well-reproduced in the present framework, in contrast to
the near-zero current for the diffusion-less system of Eq. \ref{eq:eg1}. The
actual peak shape features, for instance the value of applied $E$ for the peak
current and the peak asymmetry, are likely to differ from the deterministic
theory, for the same reason given for the potential-step chronoamperometry.

    Despite the simplicity of the models shown, it is straightforward to
simulate more complex systems, as well as increase the complexity of the
simulations by sweeping the applied parameter and sampling the associated
response. The limitations of the algorithm restricts the current range of
application, in particular for time-dependent parameter (e.g. cyclic
voltammetry). However it should be possible to simulate a wide variety of
surface reactions and setups, without having to resort to approximations or
analogies to circuit models. 

\section{Summary}
\label{sec:sum}

    The use of the Stochastic Simulation Algorithm (SSA)\cite{gillespie76} has
been explored for interfacial systems, particularly electrochemical ones.
Following Gillespie's original work\cite{gillespie92}, the Chemical Master
Equation (CME) has been obtained for such systems. The assumptions needed have
been discussed, namely the low-energy barrier between surface adsorption sites
and the use of translational as well as vibrational and rotational degrees of
freedom for thermalization of the system. The role of the electrochemical
potential, or more specifically the electrode potential, has been shown to
satisfy the requirements for the CME, as long as it remains constant in the
time interval between reactions.

    Having a CME describing the system, the SSA being deemed a logically
equivalent\cite{gillespie92}, it is assumed the possibility of its application
to simulate a time trajectory in population space. It is shown that, aside from
the associated electric response, the algorithm proceeds exactly as originally
proposed\cite{gillespie76}. For electrochemical systems, potentiostatic and
galvanostatic setups are described, and equations for each setup's electric
driving force and response are given based on an ensemble argument. This
argument is backed by the macroscopic limit of the CME/SSA
approach\cite{gillespie09b}, where the thermodynamic limit of a Langevin-type
equation (CLE)\cite{gillespie00}, together with Faraday's law of
electrolysis\cite{bard01}, is shown to result in the proposed equations.

    Finally, it is shown the results of using the SSA for electrochemical
systems, with the electrical response being given by the Butler-Volmer model
of electrochemical kinetics\cite{bard01}. A simple redox system is used for
examples, with and without a diffusion-like process. The usual, well-described
differences between the two systems are pointed out for potentiostatic setups,
together with galvanostatic results. It is shown that the current framework
reproduces the expected behaviors, with the major advantage of enabling a
rigorous quantitative comparison with experimental results through statistical
analysis, for both current and potential-controlled techniques.

    The conditions assumed and the requirements needed for the derivation of
the CME constraints the algorithm for constant current and potential-controlled
systems. The implementation of the approximations and methods developed for
the original SSA should allow the study of more complex electrochemical
systems, including both time-dependent current and potential control and
diffusion processes. It is expected that, in this way, the whole behavior of
the so-called electric double layer can be simulated using the framework
presented and its future improvements.

\begin{acknowledgments}
    The author acknowledges a Ph.D. scholarship, grant \#2013/11316-9, from
São Paulo Research Foundation (FAPESP), under the supervision of Dr. Joelma
Perez. The author is grateful to Mr. Eduardo Machado, Dr. Manuel Cruz and Dr.
Julia van Drunen for suggestions and careful examination of the manuscripts.
\end{acknowledgments}

\bibliographystyle{unsrt}
\bibliography{refs}

\begin{thebibliography}{10}

\bibitem{maioli16}
M.~Maioli, G.~Varadi, R.~Kurdi, L.~Caglioti, and G.~Pályi.
\newblock Limits of the classical concept of concentration.
\newblock {\em J. Phys. Chem. B.}, 120:7438--7445, 2016.

\bibitem{mcquarrie67}
D.~A. McQuarrie.
\newblock Stochastic approach to chemical kinetics.
\newblock {\em J. Appl. Prob.}, 4:413--478, 1967.

\bibitem{gillespie76}
D.~T. Gillespie.
\newblock A general method for numerically simulating the stochastic time
  evolution of coupled chemical reactions.
\newblock {\em J. Comput. Phys.}, 22:403--434, 1976.

\bibitem{gillespie92}
D.~T. Gillespie.
\newblock A rigorous derivation of the chemical master equation.
\newblock {\em Physica A}, 188:404--425, 1992.

\bibitem{gillespie77}
D.~T. Gillespie.
\newblock Exact stochastic simulation of coupled chemical reactions.
\newblock {\em J. Phys. Chem.}, 81:2340--2361, 1977.

\bibitem{gillespie13}
D.~T. Gillespie, A.~Hellander, and L.~R. Petzold.
\newblock Perspective: stochastic algorithms for chemical kinetics.
\newblock {\em J. Chem. Phys.}, 138:170901, 2013.

\bibitem{gillespie01}
D.~T. Gillespie.
\newblock Approximate accelerated stochastic simulation of chemically reacting
  systems.
\newblock {\em J. Chem. Phys.}, 115:1716, 2001.

\bibitem{gillespie07}
D.~T. Gillespie.
\newblock Stochastic simulation of chemical kinetics.
\newblock {\em Annu. Rev. Phys. Chem.}, 58:35--55, 2007.

\bibitem{gillespie09a}
D.~T. Gillespie.
\newblock A diffusional bimolecular propensity function.
\newblock {\em J. Chem. Phys.}, 131:164109, 2009.

\bibitem{gillespie00}
D.~T. Gillespie.
\newblock The chemical {Langevin} equation.
\newblock {\em J. Chem. Phys.}, 113:297, 2000.

\bibitem{gillespie09b}
D.~T. Gillespie.
\newblock Deterministic limit of stochastical chemical kinetics.
\newblock {\em J. Phys. Chem. B}, 113:1640--1644, 2009.

\bibitem{bard96}
A.~J. Bard and F-R.~F. Fan.
\newblock Electrochemical detection of single molecules.
\newblock {\em Acc. Chem. Res.}, 29:572--578, 1996.

\bibitem{dick15}
J.~E. Dick and A.~J. Bard.
\newblock Recognizing single collisions of {PtCl$_6^{2-}$} at femtomolar
  concentrations on ultramicroelectrodes by nucleating electrocatalytic
  clusters.
\newblock {\em J. Am. Chem. Soc.}, 137:13752--13755, 2015.

\bibitem{kim16}
J.~Kim and A.~J. Bard.
\newblock Electrodeposition of single nanometer-size {Pt} nanoparticles at a
  tunneling ultramicroelectrode and determination of fast heterogeneous
  kinetics for {Ru(NH$_3$)$_6^{3+}$} reduction.
\newblock {\em J. Am. Chem. Soc.}, 138:975--979, 2016.

\bibitem{heer71}
C.~V. Heer.
\newblock Statistical thermodynamic theory for adsorption isotherms.
\newblock {\em J, Chem. Phys.}, page 4066, 1971.

\bibitem{bard01}
A.~J. Bard and L.~R. Faulkner.
\newblock {\em Electrochemical Methods: Fundamentals and Applications}.
\newblock John Wiley \& Sons, 2nd edition, 2001.

\bibitem{Note1}
The MATLAB script and a Python3 version are available in
  https://github.com/beruski/ec-ssa.

\bibitem{bockris02}
J.~O'M. Bockris, A.~K.~N. Reddy, and M.~Gamboa-Aldeco.
\newblock {\em Modern Electrochemistry}, volume~2A.
\newblock Kluwer Academic, 2nd edition, 2002.

\end{thebibliography}

\end{document}